\documentclass[usenatbib]{mnras}
\usepackage{newtxtext,newtxmath}
\usepackage[T1]{fontenc}
\usepackage{ae,aecompl}
\usepackage[english]{babel}
\usepackage{graphicx}
\usepackage[dvipsnames,svgnames,x11names]{xcolor}
\usepackage{tikz}
\usepackage{siunitx}
\usepackage{multirow}
\usepackage{soul}
\usepackage{fancyvrb}
\usepackage{multicol}
\usepackage{adjustbox}
\usepackage{verbatim}
\usepackage{float}
\usepackage{amsmath}
\usepackage{amssymb}
\usepackage{pifont}
\usepackage{soul}
\newcommand{\cmark}{\ding{51}}

\newcommand*\mathinhead[2]{\texorpdfstring{$\boldsymbol{#1}$}{#2}}
\definecolor{scc}{rgb}{0.0, 0.26, 0.15}
\definecolor{cadmiumred}{rgb}{0.89, 0.0, 0.13}

\title[SFR estimation with Machine Learning]{Star Formation Rates for photometric samples of galaxies using machine learning methods}

\author[Delli Veneri et al.]{M.~Delli~Veneri$^{1}$\thanks{E-mail:
micheledelliveneri@gmail.com},  S. Cavuoti$^{1,2,3}$\thanks{E-mail:
stefano.cavuoti@gmail.com}, M. Brescia$^{1}$, G. Longo$^{2,3}$,  G. Riccio$^{1}$\\
$^{1}$INAF - Astronomical Observatory of Capodimonte, via Moiariello 16, I-80131, Napoli, Italy\\
$^{2}$Department of Physics ``E. Pancini'', University Federico II, via Cinthia 6, I-80126, Napoli, Italy\\
$^{3}$INFN section of Naples, via Cinthia 6, I-80126, Napoli, Italy}
\date{Accepted 2019 March 19. Received 2019 March 19; in original form 2018 November 16}
\pagerange{\pageref{firstpage}--\pageref{lastpage}} \pubyear{2019}

\begin{document}

\label{firstpage}
\maketitle

\begin{abstract}
Star Formation Rates or SFRs are crucial to constrain theories of galaxy formation and evolution. SFRs are usually estimated via spectroscopic observations requiring large amounts of telescope time. We explore an alternative approach based on the photometric estimation of global SFRs for large samples of galaxies, by using methods such as automatic parameter space optimisation, and supervised Machine Learning models. We demonstrate that, with such approach, accurate multi-band photometry allows to estimate reliable SFRs. We also investigate how the use of photometric rather than spectroscopic redshifts, affects the accuracy of derived global SFRs.
Finally, we provide a publicly available catalogue of SFRs for more than 27 million galaxies extracted from the Sloan Digital Sky survey Data Release 7. The catalogue is available through the Vizier facility at the following link \url{ftp://cdsarc.u-strasbg.fr/pub/cats/J/MNRAS/486/1377}.

\end{abstract}

\begin{keywords}
techniques: photometric - galaxies: distances and redshifts - galaxies: photometry - methods: data analysis - catalogues 
\end{keywords}

%%%%%%%%%%%%%%%%%%%%%%%%%%%%%%%%%%%%%%%%%%%%%%%%%%%%%%%%%%%%%%%%%%%%%%%%%%%%%%%%%%%%%%
\section{Introduction}\label{SEC:introduction}

During the last few year, multi-wavelength surveys have led to a remarkable progress in producing large galaxy samples that span a huge variety of galaxy properties and redshift. All together, these data provided us with reliable information for many hundred thousand galaxies \citep{DR7, Salvato2009, Salvato2011, Marchesi2016, Cardamone2010, Matute2012} and have triggered similar improvements in the determination of physical parameters crucial to  understand and constrain galaxy formation and evolution. 
Among these parameters, the global Star Formation Rate or SFR \citep{madau}, provides a luminosity-weighted average across local variations in star formation history and physical conditions within a given galaxy. 

Broadly speaking, SFR estimators are usually derived from measured fluxes, either monochromatic or integrated over some specific wavelength ranges, selected in order to be sensitive to the short-lived massive stars present in a given galaxy. 
In the literature, there is a large variety of such estimators spanning from the UV/optical/near-IR range ($\sim$ 0.1 - 5 $\mu m$), which probes the stellar light emerging from young stars, to the mid/far-IR ( $\sim$ 5 - 1000 $\mu m$), which instead probes the stellar light reprocessed by dust \citep{Kennicut&Evans,Kennicutt1998}. Other estimators rely on the gas ionized by massive stars \citep{calzetti2004,hong2011},  hydrogen recombination lines, forbidden metal lines, and in the millimeter range, the free-free (Bremsstrahlung) emission \citep{Schleicher2013}. Finally, other estimators can, at least in principle, be derived in the X-ray domain, from X-ray binaries, massive stars, and supernovae via the non-thermal synchrotron emission, following early suggestions by \cite{condon1992}. \\
An ample literature, however, shows that the correct derivation of SFRs from optical/FIR broad band data is a highly non-trivial task, due to the complex  and still poorly understood correlation existing between the SFR and the broad band photometric properties integrated over a whole galaxy \citep{Pearson2018,Cooke2018,Fogarty2017,Rafelski2016}.\\
Each estimator is sensible to a specific and different SFR timescale and thus a proper understanding of the SFR phenomenology requires a combination of different estimators; in particular, UV and total IR radiations are sensible to the longer timescales, $ \sim 10^8 yr$, while the ionising radiation is sensitive to the shortest timescales, $ \sim 10^6 yr$. Furthermore, optical and UV estimators often need corrections to account for dust presence and, for this reason, they are not used on their own, but in combination with other estimators \citep{Calzetti2007}.
Another methodology suitable to estimate SFRs for large samples of objects is the so-called spectral energy distribution (SED) template fitting, which compares an observed galaxy spectrum with a large database of template spectra, generated by stellar population synthesis models \citep{Conroy2013}. This method, however, suffers from the age-dust-metallicity degeneracy and, in order to reliably measure ages and hence SFRs, high quality data are required and, due to the choice of template spectra, severe biases are often introduced in the resulting ages. 
In a seminal paper \cite{Wuyts2011}, SFRs for galaxies at $z_{spec} \sim 3$ were derived using all the methods previously explained, finding that all estimators agree with no systematic offset, providing that an extra attenuation toward $H_{II}$ regions is included when modelling the $H\alpha$ SFRs. Nevertheless, the same paper also concluded that, at high redshift, nebular emission lines may introduce a systematic uncertainty affecting the derived specific SFRs by a factor of two.
The present work takes place in the framework of the new discipline of Astroinformatics, which aims at allowing the scientific exploitation of large data sets produced by the modern digital, panchromatic and multi-epoch surveys, using a variety of techniques largely derived from, but not restricted to, the statistical learning domain. In this framework a new viable approach to obtain SFR estimates for large samples of objects was recently presented by  \cite{Stensbo}, who transformed the SFR estimation into a machine learning (ML) non-linear regression problem. With this method, the only prerequisite is the availability of a sufficient amount of objects with well measured SFRs, to be used as the training/validation sample. We follow a similar approach and use exactly the same data in order to compare our results with those in \cite{Stensbo}.
A parallel and independent machine learning approach was used in \cite{Bonjean} to solve the SFR regression problem with 
three main differences with respect to our approach: 1) they use shallow-IR instead of our optical features, 2) they employ a classical feature selection technique (embedded in their Random Forest model), and 3) they include spectroscopic information into the training parameter space. In particular, we investigate how effective ML based methods can be in deriving SFRs in large samples of galaxies, paying special attention to \textit{feature selection}, i.e. to the selection of the most suitable parameter space. As we shall demonstrate, the selection of the optimal set of features, in addition to a more accurate prediction, can also be used to derive an insight into the physics of the phenomenon \citep{brescia:2018b}.

 In Sec.~\ref{SEC:data} we introduce the data and in  Sec.~\ref{SEC:Algorithms} all algorithms and ML methods used. In Sec.~\ref{sec:exp} we describe our campaign of experiments and related results. Finally, in Sec.~\ref{sec:concl} we discuss the results and draw some conclusions.
 
\section{Data}\label{SEC:data}
Since we were also interested in comparing our results with those presented in \cite{Stensbo}, the same data, derived from the Sloan Digital Sky Survey Data Release $7$ (SDSS-DR7), have been used \citep{DR7}. 
Such data release has also been used by \cite{Brinchmann} to derive reliable SFRs for a subsample of $\sim 10^6$ galaxies, through  a full analysis of the emission and absorption line spectroscopy, available in the SDSS spectroscopic data set (hence not based on the $H_{\alpha}$ flux alone). The reliability of this study was confirmed in \cite{Salim2007}, who carried out an independent study using optical photometry from the SDSS and near UV measurements from GALEX, thus bypassing some uncertainties inherent the spectroscopic $H_{\alpha}$ aperture corrections. The local SFRs (normalized to $z = 0.1$) from the two studies (\citealt{Brinchmann,Salim2007}) turned out to agree within the errors.

The final catalogue contains several types of magnitudes\footnote{\url{http://classic.sdss.org/dr7/algorithms/photometry.html}}: \textit{psfMag}, \textit{fiberMag}, \textit{petroMag}, \textit{modelMag}, \textit{expMag} and \textit{deVMag} in the u, g, r, i, and z bands; it includes also the \textit{spectroscopic redshift} ($z_{spec}$), the \textit{photometric redshift} ($photoz$), derived using an hybrid combination of a template fitting approach with an empirical calibration using objects with both observed colours and spectroscopic redshift \citep{Csabai2007}, as well as the \textit{Average Specific Star Formation Rate} (hereafter SFR). Starting from this dataset, we performed a pre-processing, in which the following constrains were applied to improve the reliability of the final knowledge base:

\begin{enumerate}
\item we required high quality estimations of \textit{SFR}, i.e. objects for which the quality flag is  equal to $0$ (see  \citealt{Brinchmann} for further details);
\item we required high quality spectroscopic redshifts (i.e. with \textit{zWarning} = 0; see \citealt{DR7} for further details);
\item all objects affected by missing information, namely objects with at least one feature having a ``\textit{Null value}'', were removed from the knowledge base, since our chosen ML methods are not capable of handling missing features.
\end{enumerate}

\begin{figure*}
\centering
\includegraphics[width=\textwidth]{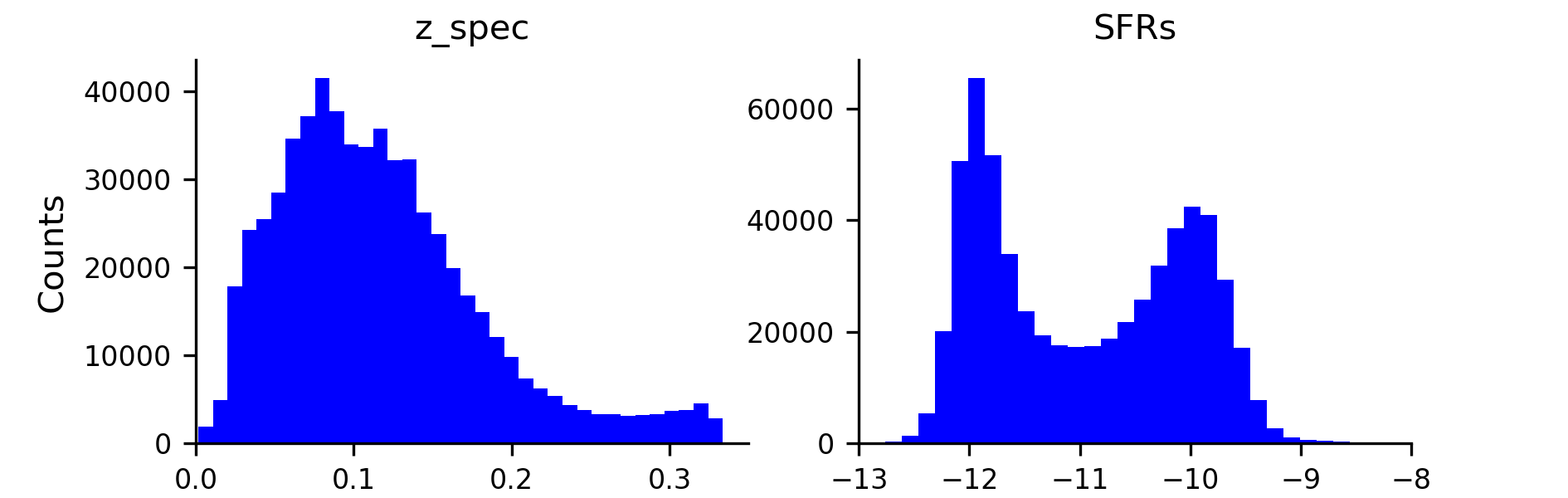}
\vspace*{-6mm}
\caption{Spectroscopic redshift (left panel) and SFR (right panel) distributions of the knowledge base.} \label{fig:specz}
\end{figure*}

 The final knowledge base consists of $603,680$ galaxies, respectively, $362,208$ for training and $241,472$ as blind test set, extracted through a random shuffling and split procedure. Furthermore, for each magnitude type we derived the related colours, i.e. \textit{u-g}, \textit{g-r}, \textit{r-i}, and \textit{i-z}, thus reaching a total of $56$ features, $55$ photometric (magnitudes, colours and $photoz$) and one spectroscopic ($z_{spec}$). Finally, we added the SFR, used as target variable. The distribution of spectroscopic redshifts and SFRs for the knowledge base is shown in Fig.~\ref{fig:specz}. 

%%%%%%%%%%%%%%%%%%%%%%%%%%%%%%%%%%%%%%%%%%%%%%%%%%%%%%%%%%%%%%%%%%%%%%%%%%%%%%%%%%%%%%
\section{The methods}\label{SEC:Algorithms}
In the present work we make use of two supervised Machine Learning (ML) methods:
%, respectively: 
Random Forest (RF, \citealt{Breiman20015}) and Multi Layer Perceptron trained by the Quasi Newton Algorithm (MLPQNA, \citealt{Brescia20121155}).
Furthermore, in order to optimise their performances, we apply k-fold cross validation (cf. \citealt{Kohavi95astudy}) and a novel feature selection model called \textit{Parameter handling investigation LABoratory} ($\Phi$LAB, \citealt{brescia:2018}). These methods are shortly described in the following sections. 

\subsection{Random Forest}
The Random Forest \citep{Breiman20015} operates by generating an ensemble of decision trees during the training phase, based on different subsets of input data samples. 
%Within the tree construction, different combinations of features internal to data patterns are included in the decision chain. 
For each decision tree, a random subset of input features is selected and used to build the tree.
By imposing a sufficient number of trees (depending on the parameter space complexity and input data amount), all given features will, with high probability, be examined within the produced forest \citep{hastie}.
%The Random Forest algorithm  used in our experiments has been implemented with the Python scikit-learn library \citep{scikit-learn}.
In our experiments, we make use of the random forest implementation from the {\small PYTHON} library scikit-learn \citep{scikit-learn}.
For our purposes we heuristically choose an ensemble of $1000$ trees, trying to reach a good trade-off between performance and training computing time. Each tree was created by a random shuffling of the full set of features available and with a minimum split at each node equal to two.

\subsection{MLPQNA}
The Multi Layer Perceptron trained by the Quasi Newton Algorithm (MLPQNA) is a model in which the learning rule is based on the Quasi Newton rule, one of the Newton's methods aimed at finding the stationary point of a function and based on an approximation of the Hessian of the training error through a cyclic gradient calculation. MLPQNA makes use of the known L-BFGS algorithm (Limited memory - Broyden Fletcher Goldfarb Shanno, \citealt{byrd1994}). Our multilayer perceptron architecture consists of two hidden layers with respectively $2N + 1$ and $N - 1$ neurons, where $N$ is the number of input features. %\ks{Question: is the second layer really $N-1$ neurons, or should it have been $N+1$?}
All further details of the MLPQNA implementation, as well as its performance in different astrophysical contexts, have been extensively discussed elsewhere \citep{Brescia20121155,Brescia2014,Brescia20153893,Cavuoti2013968,Cavuoti201545,Cavuoti20171959,DIsanto20163119}. With respect to the Random Forest, our actual implementation of the MLPQNA model is generally more computationally intensive and thus some of the experiments performed later on in this paper are referred to the Random Forest model only.
\subsection{\textit{K}-fold Cross-Validation}
Within the context of the supervised machine learning paradigm, it is common practice to exploit the available knowledge base by deriving three disjoint subsets: one (training set) to be used for learning purposes, namely to acquire the hidden correlation among input features and the output target; a second (validation set) to check the training status, in particular, to measure the learning level and to verify the absence of any loss of generalisation capabilities (a phenomenon also known as overfitting); and the third one, the test set, is used to evaluate the overall performance of the trained and validated model. The latter two datasets are blind or, in other words, they do not contain input patterns already used during the training phase \citep{Brescia2013}.

In some cases, especially in presence of a limited amount of samples available within the knowledge base, a valid alternative approach, also applied in this work, is the so called \textit{k-fold} cross-validation technique \citep{Kohavi95astudy}. This is an automatic cross-validation procedure, based on \textit{k} different training sessions, specified as it follows: (\textit{i}) random splitting of the training set into \textit{k} random subsets, each one composed by the same fraction of the knowledge base; (\textit{ii}) each of the $k$ subsets is then, in turn, used as test set, while the remaining $k-1$ subsets are used for training/validation.

The purpose of $k$-fold cross-validation is, in part, to test the model's performance stability on different subsets of the data, thus making sure that a chosen training/test set was neither particular favourable or unfavourable, and to minimise the risk of any training overfitting occurrence.
In our case we heuristically choose $k = 10$, representing a good compromise between computing efficiency and data amount within the folds.

\subsection{Feature selection} \label{sec:fs}
Not all input features contain the same amount of information for a particular problem domain, and discovering the most informative variables may, on the one hand, drastically reduce the computing time and, on the other, it can provide useful insights into the physical nature of the problem. In this work we used a novel feature selection method, called \emph{Parameter handling investigation LABoratory} \citep[$\Phi$LAB,][]{brescia:2018}.

The choice of an optimal set of features is connected to the concept of \textit{feature importance}, based on the measure of a feature's \textit{relevance}. Formally, the importance of a feature is its percentage of informative contribution to a learning system.

We approach the feature selection task on two complexity levels: (\textit{a}) the \textit{minimal-optimal feature selection}, which consists of a selection of the smallest parameter space able to obtain the best learning performance; and (\textit{b}) the \textit{all-relevant feature selection}, able to extract the most complete parameter space, i.e. all features considered relevant for the solution to the problem. The second level is appropriate for problems with highly correlated features, as these features will contain nearly the same information. With a minimal-optimal feature selection, choosing any one of them (which could happen at random if they are perfectly correlated) means that the rest will never be selected.

%evidently more complex, in particular to minimise the occurrence of the high-correlation compromise issue, i.e. it turns out to be extremely helpful in the frequent cases of highly correlated features, where it is difficult to evaluate the relevance contribution of individual features. In such cases the commonly adopted strategy is to equally partition their importance by associating all such features to the same relevance class. This is evidently a simplification that could cause useless redundancy in the parameter space.

We investigated the possibility to find a method able to optimise the parameter space, by solving the all-relevant feature selection problem, thus indirectly improving the physical knowledge about the problem domain. The method presented, $\Phi$LAB, includes properties of both embedded and wrappers categories of feature selection \citep[see][for an introduction to feature selection]{Guyon2003}. The details of the method are presented in the Appendix \ref{sec:philabmethod}.

\subsection{Evaluation Metrics}
In order to evaluate the performance of our experiments
we use the quantity $\Delta_{SFR}$, defined as:
$$ \Delta_{SFR} \equiv SFR_\text{photometric}- SFR_\text{spectroscopic}$$
where $SFR_\text{photometric}$ is the estimated SFR, $SFR_\text{spectroscopic}$ is the target value obtained from spectroscopy. We indicate also $S_{m}$ as the blind test set. Then we use the following metrics:

\begin{itemize}
\item $RMSE = \sqrt{\frac{1}{|S_{m}|} \sum_{n \in S_{m}} [\Delta_{SFR}]^{2}}$, the root-mean-square error of the residuals;
\item  $ Median(\Delta_{SFR})$, the median of the residuals;
\item $\sigma = \sqrt{\frac{1}{|S_{m} - 1|} \sum_{n \in S_{m}} [\Delta_{SFR} - \overline{\Delta_{SFR}}]^{2}}$, the standard deviation of the residuals;
\item $\eta$, the percentage of catastrophic outliers. According to the definition by \cite{Stensbo}, we consider an outlier to be catastrophic if $\Delta_{SFR} > 3 \sigma$. %We computed the fraction of catastrophic outliers $\eta$ for each experiment. 
Consequently, the percentage of outliers depends on the value of  $\sigma$.
\end{itemize}

The RMSE and $\sigma$ turned out to be almost identical in all of our experiments; the mathematical relation between the two estimators is: $RMSE =  \sqrt{ \big( \overline{\Delta_{SFR}}^2 + \sigma^2 \big) }$. This means that the mean of $\Delta_{SFR}$  is negligible. We decided to report only the  RMSE in each table. Nevertheless, we will use both estimators since the RMSE is used to evaluate the model performance, while the $\sigma$ is used to compute the fraction of catastrophic outliers.

\section{Experiments and Results}\label{sec:exp}
In order to optimize the procedure in terms of SFR accuracy, we performed a series of experiments.

As first step we evaluated the performance of our regression models on the entire set of available features. 
Afterwards we evaluated the usefulness of the k-fold cross validation, by verifying if such time-consuming operation (in our case it extends the training time of the network by almost a factor of ten) is effectively required to minimize overfitting and to check how the models perform on different datasets. In other words, how stable are the results across the whole datasets.
Subsequently, we performed a feature selection to optimise the parameter space, indirectly suitable also for a comparison with the feature selection described in \cite{Stensbo}. Then we performed a series of experiments to evaluate the most appropriate size of the training set.
After that we analysed the relationship between the photometric redshift quality and the accuracy of SFRs. Finally, we compared the SFR prediction performance between the methods RF and MLPQNA on the best set of features found by $\Phi$LAB.

\subsection{RF and MLPQNA performances on the full set of photometric features} \label{sec:RF_VS_MLPQNA_54}
As said above, we performed a preliminary performance test using the full set of available features (i.e. the $54$ photometric features described in Sec. \ref{SEC:data}). The results are summarised in Table~\ref{tab:RF_VS_MLPQNA_54}.

\begin{table}
    \centering
    \begin{tabular}{| c | c | c | c |}
         \hline
         \bf Model & \bf RMSE & \bf Median &  {$\boldsymbol{\eta}$} \\
         \hline
         RF & 0.252 & -0.021 &  1.99 \\
         MLPQNA & 0.261 & -0.016  & 1.76 \\
         \hline
    \end{tabular}
    \caption{Performance comparison of the RF and MLPQNA models, calculated on the blind test set, using all the $54$ photometric features available and the full training set.}
    \label{tab:RF_VS_MLPQNA_54}
\end{table}
The results of Table~\ref{tab:RF_VS_MLPQNA_54} show that RF performs better than MLPQNA. 

\subsection{k-fold Cross Validation} \label{sec: k-fold results}
%\ks{As per above, this entire section is a misunderstanding and needs to go, and the uncertainties need to be added back into the tables. Sorry, but this is very important to me.}
As a preliminary step of the training phase, and accordingly to what was done in \cite{Stensbo}, we decided to verify if the k-fold cross validation technique is required to avoid overfitting in this particular use-case. Therefore, we replicated the RF and MLPQNA performance tests on the full set of $54$ photometric features (see Sec.~\ref{sec:RF_VS_MLPQNA_54}), but this time implementing the k-fold cross validation using $k=10$. \\
The results of the experiment can be seen in Table~\ref{tab:cv} where we compare the RF and MLPQNA performances with and without k-fold cross validation. Experiments with cross validation, while increasing the computing time by almost an order of magnitude, do not show any significant improvement in terms of accuracy. In Table~\ref{tab:cv_errors} the standard deviations of the used statistical estimators computed over the ten folds are shown. As it can be seen, the results show that the cross-validation contribution is negligible, thus confirming that the information in the Knowledge Base is well distributed and, as a consequence, that both models are capable to work in a stable way across different datasets, as well as the fact that they are intrinsically robust against overfitting.
For such reasons we decided to perform all further experiments without the k-fold cross validation technique.
%\ks{You shouldn't use training time as an argument for anything, as it is both implementation and hardware specific. Using a modern library, e.g. keras, a full 10 fold CV of a neural network with the mentioned architecture (113-57-1 neurons) should take less than a day on a laptop.}

\begin{table}
    \centering
    \setlength\tabcolsep{3pt}\resizebox{\columnwidth}{!}{\centering
    \begin{tabular}{| c | c | c | c | c | c | c |}
    \hline
     \multirow{2}{*}{\bf Model} & \multicolumn{3}{c|}{cross-validation}&\multicolumn{3}{c|}{no cross-validation}\\ \cline{2-7}
     \rule{0pt}{3ex}
      & {\bf RMSE} & {\bf Median} &  {$\boldsymbol{\eta}$} & {\bf RMSE} & {\bf Median}  & {$\boldsymbol{\eta}$}\\
     \hline 
     \rule{0pt}{3ex}
        RF & 0.252 & -0.021 &  1.99 & 0.252 & -0.021 &  2.07\\
        MLPQNA & 0.261 & -0.016 &  1.76 & 0.261 & -0.016 & 1.78 \\
    \hline
    \end{tabular}}
    \caption{Experiments result with and without k-fold cross validation. The statistics are calculated on the blind test set only.}
    \label{tab:cv}
\end{table}

\begin{table}
    \centering
    \begin{tabular}{| c | c | c | c | c |}
        \hline
         \textbf{Model} & $\sigma_{RMSE}$ & $\sigma_{Median}$ & $\sigma_{\sigma}$ & $\sigma_{\eta}$ \\
         \hline
         RF & $0.001$ & $0.00003$ & $0.001$ & $0.041$ \\
         MLPQNA & $0.002$ & $0.00051$ & $0.002$ & $0.002$ \\
         \hline
    \end{tabular}
    \caption{Effect of the cross-validation on the experiments of Table \ref{tab:cv}. Each column represents the standard deviation across ten different experiments for a statistical estimator. In this case our spectroscopic SFRs span in the range $\sim$ ]$-14$, $-17$[. This shows how, in this case, the cross-validation can be considered as negligible.}
    \label{tab:cv_errors}
\end{table}

\subsection{Feature Selection results} \label{sec:fs_results}
\begin{figure}
\centering
\includegraphics[width=\columnwidth]{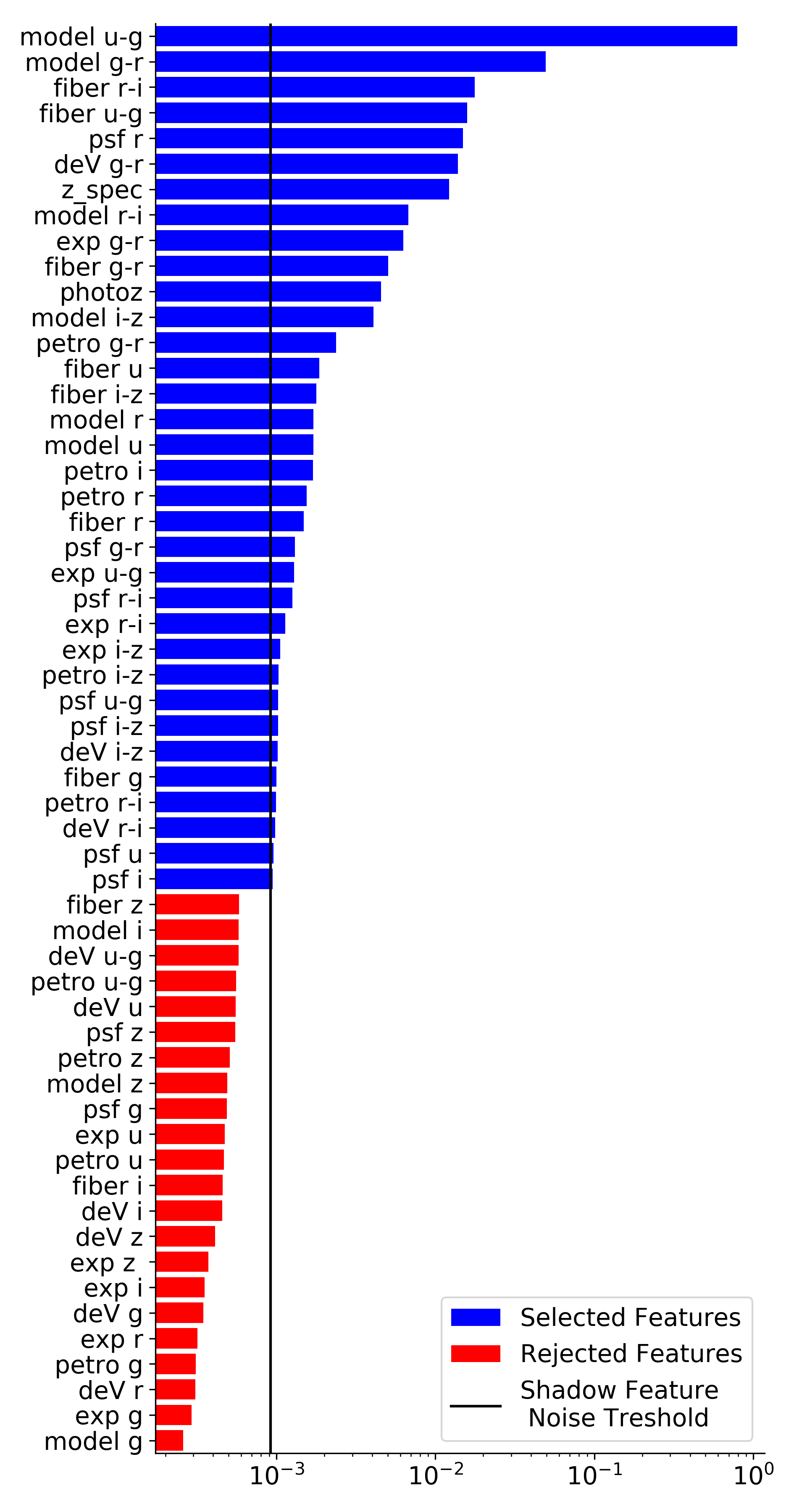}
\caption{Feature Importance percentages derived by applying the $\Phi$LAB method to the full knowledge base and parameter space available, described in Sec.~\protect\ref{SEC:data}. In blue  are marked the selected features, while in red those rejected by the method. The vertical black line is the noise threshold computed through the \textit{shadow feature} technique embedded in the $\Phi$LAB algorithm (see Sec.~\protect\ref{sec:fs} for details). The noise threshold corresponds to an importance value of $\sim 0.062\%$. }
\label{fig:featImp}
\end{figure}

To perform the feature selection, We made use of our model $\Phi$LAB using the full knowledge base available (see Sec.~\ref{SEC:data}). The $34$ features selected by the method are shown in Fig.~\ref{fig:featImp} and listed in Table~\ref{tab:features}.

\begin{table}
\resizebox{\columnwidth}{!}{\centering
\begin{tabular}{|c | c c c c c c|}
\hline
{\bf{Feature}} & {\bf model} & {\bf fiber} & {\bf psf} & {\bf exp} & {\bf petro} & {\bf deV} \\
\hline \rule{0pt}{3ex}%  EXTRA vertical height
\textbf{u-g} & \cmark &  \cmark &  \cmark &  \cmark & & \\
\textbf{g-r} &  \cmark &  \cmark &  \cmark &  \cmark &  \cmark &  \cmark \\
\textbf{r-i} &  \cmark &  \cmark &  \cmark &  \cmark &  \cmark &  \cmark \\
\textbf{i-z} &  \cmark &  \cmark &  \cmark &  \cmark &  \cmark &  \cmark \\
\textbf{u}   &  \cmark &  \cmark &  \cmark & & & \\
\textbf{g}   & &  \cmark & & & & \\
\textbf{r}   &  \cmark &  \cmark &  \cmark & &  \cmark & \\
\textbf{i}   & & &  \cmark & &  \cmark & \\
\hline \hline
& \multicolumn{3}{c}{$\boldsymbol{z_{spec}}$} & \multicolumn{3}{c|}{$\boldsymbol{photoz}$} \\
\hline \rule{0pt}{3ex}%  EXTRA vertical height
\textbf{redshift}& \multicolumn{3}{c}{\cmark} & \multicolumn{3}{c|}{\cmark} \\
\hline
\end{tabular}
}\caption{List of features selected by $\Phi$LAB running on the full knowledge base available.}
\label{tab:features}
\end{table}

Concerning the excluded features, as it can be seen from Fig.~\ref{fig:featImp}, $\Phi$LAB marks as ``unimportant'' all the $z$ band magnitudes, five out of six \textit{g} band magnitudes (retaining only \textit{fiberMag\_g} but with a very low ranking), three out of six \textit{u} band magnitudes, four out of six \textit{i} band magnitudes and two in the \textit{r} band. Conversely, all colours were retained with the exception of \textit{devMag\_u-g} and \textit{petroMag\_u-g}.

In particular, from Fig.~\ref{fig:featImp}, we can notice that all the exponential and de Vaucouleurs magnitudes are excluded (while their colours are retained) in favour of the \textit{modelMag}\footnote{\url{http://classic.sdss.org/dr7/algorithms/photometry.html}}. For the other types of magnitudes only two or three are dropped (i and z for \textit{fiberMag}, g, z and i for \textit{modelMag}, u, g and z for the \textit{petroMag} and g, i and z for the \textit{psfMag}). All together this leads to a total of $22$ rejected features.\\

\begin{figure}
\centering
\includegraphics[width=\columnwidth]{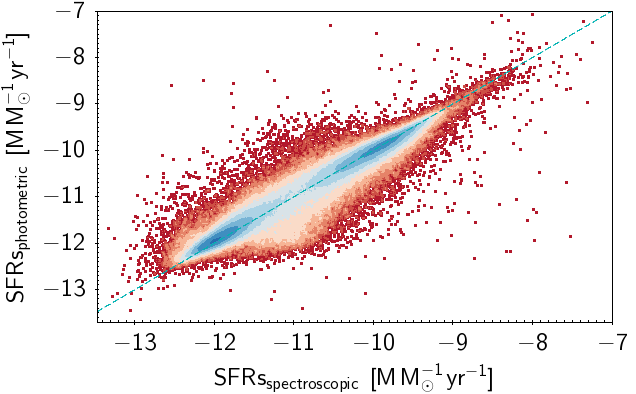}
\caption{$SFRs_{spectroscopic}$ VS $SFRs_{photometric}$ scatter plot related to the MLPQNA$_{\Phi LAB}$ experiment, selected to produce the final SFR catalogue (see Appendix~\ref{catalogue}).} \label{fig:specvspredict} 
\end{figure}

The optimised parameter space identified by $\Phi$LAB (i.e. the $32$ selected features of Fig.~\ref{fig:featImp}, excluding the two redshifts) was employed to perform a comparison between the two machine learning regression models used to estimate the SFR, starting from the same knowledge base. Table~\ref{tab:mlpqnavsrf} reports the results, while the distribution of photometric vs spectroscopic SFRs for MLPQNA is shown in Fig.~\ref{fig:specvspredict}. The MLPQNA obtains the best performance, ($\sim 1.5\%$ better accuracy than the RF on the same data). However, this comes at the cost of a much higher computational time, 
since using $32$ features the RF takes $\sim 0.05\%$ of the computational time required by MLPQNA and this ratio further decreases for an increasing number of features.
In spite of this, we decided to use the MLPQNA model to produce the SFRs catalogue presented in Appendix \ref{catalogue}.

In principle, a robust feature selection method should be able to identify the most relevant features in a way as independent as possible from the specific machine learning model used to subsequently approach the regression problem. Furthermore, in order to verify that the selected feature space is the best choice, a supplementary set of regression performance tests should be performed by using alternative subsets of features. In what follows we discuss these two aspects.

\begin{figure}
    \centering
    \includegraphics[width=\columnwidth]{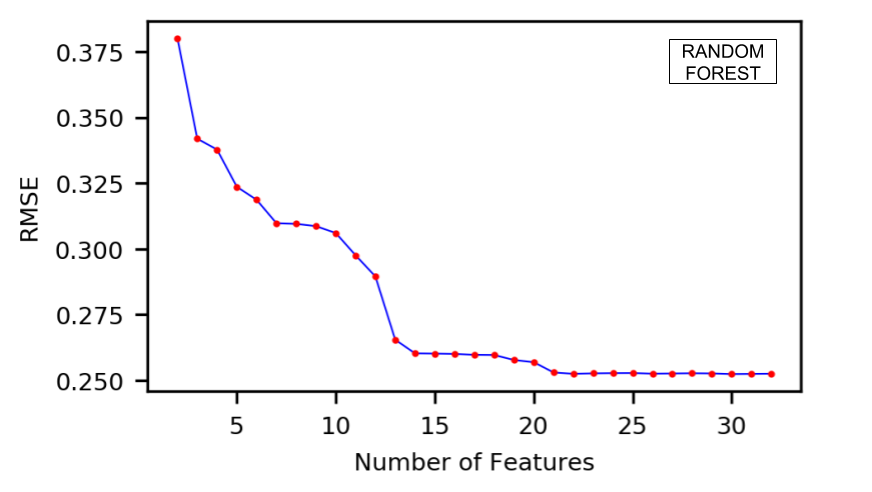}
    \includegraphics[width=\columnwidth]{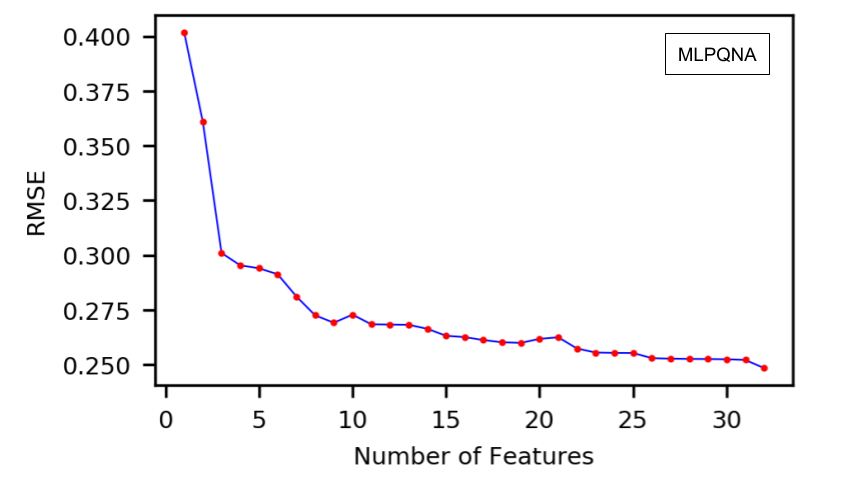}
    \caption{Performance variation of the Random Forest (upper panel) and MLPQNA (lower panel) models with respect to the number of features used in the training. On the y-axis we report the RMSE value computed on the blind test set, while on the x-axis the incremental number of features included in the training.}
    \label{fig:RMSE_VS_FEATURES}
\end{figure}

In order to verify the independence of the feature selection on the two regression methods, we iteratively trained the RF and MLPQNA, using always the entire training set, starting with just one feature and adding, at each iteration, a new feature (in the order of importance selected by $\Phi$LAB). until all the $32$ photometric features selected by $\Phi$LAB were used. Fig.~\ref{fig:RMSE_VS_FEATURES} shows the RMSE as function of the number of used features for both RF and MLPQNA methods. As it can be seen, the RMSE decreases steadily with the number of features in both cases, reaching the minimum value when the all $32$ features are considered.

To further investigate the capability of the $\Phi$LAB method to identify the optimal parameter space, we performed the following additional experiments with the RF:
\begin{itemize}
    \item \textit{RND}: we performed ten experiments all using the same number of features ($32$) found by $\Phi$LAB, but randomly extracted from the original parameter space (excluding the redshifts). These experiments were performed in order to compare, fixed the number of features selected by $\Phi$LAB, the performances achieved by the best all-relevant features experiment (RF$_{\Phi LAB}$ experiment) with those obtained via a random extraction;
    \item \textit{B+W (Best plus Worst)}:
    this experiment was performed in order to confirm the lack of relevance of the rejected features and also to investigate why the method rejected some features which at least should have conveyed relevant information. Therefore we used the best $10$ features selected by $\Phi$LAB (excluding redshift) plus the $22$ features rejected by $\Phi$LAB, in order to maintain fixed to $32$ the amount of used feature. 
\end{itemize}

The results of these experiments are reported in Table~\ref{tab:Philab_Test}. The experiment reaching the best performance is RF$_{\Phi LAB}$, thus confirming the reliability of the $\Phi$LAB method in optimising the parameter space by selecting the all-relevant subset of features best suited to solve the regression problem.
Nevertheless the $\Phi$LAB and B+W experiments show a very similar performance. Such behaviour seems to indicate that most of weak relevant and rejected features bring the same amount of contribution to solve the regression problem, and that $\Phi$LAB rejects those features considered as redundant.
%\ks{Comment: The following text seems to be the logical continuation of the text just before my comment about evaluation of the features with MLPQNA. Everything in between discusses the RF, so the context moves from MLPQNA->RF->MLPQNA. I suggest you restructure the section.}
For the reasons already mentioned and related to the computational cost of MLPQNA, these experiments were performed using the RF only.
Anyway, the fact that MLPQNA using the $32$ features selected by $\Phi$LAB (MLPQNA$_{\Phi LAB}$ experiment) achieves better performances than when the entire set of $54$ photometric features is used (table~\ref{tab:RF_VS_MLPQNA_54}), indirectly confirms the reliability of the set of features selected by $\Phi$LAB.

\begin{table}

\centering
\centering
    \begin{tabular}{| l | l | l | l |}
        \hline
        
        \bf ID & {\bf RMSE} & {\bf Median} &  {$\boldsymbol{\eta}$} \\
        \hline              
        \rule{0pt}{3ex}%  EXTRA vertical height
        RF$_{\Phi LAB}$ & 0.252 & -0.021 & 2.03\\
        MLPQNA$_{\Phi LAB}$ & 0.248 & -0.017 & 1.99\\
        \hline
    \end{tabular}
    \caption{Comparison between MLPQNA and RF models using the $32$ photometric features identified by $\Phi$LAB. Both models have been applied to the same training and blind test sets.}
\label{tab:mlpqnavsrf}
\end{table}

\begin{table}
    \centering
    %\resizebox{\columnwidth}{!}{\centering
    \begin{tabular}{| c | c | c | c |}
        \hline
        \bf ID & {\bf RMSE} & {\bf Median} &  {$\boldsymbol{\eta}$} \\
        \hline              
        \rule{0pt}{3ex}%  EXTRA vertical height
        RF$_{\Phi LAB}$    & 0.252 & -0.021 & 2.03\\
        RND             & 0.269 & -0.018 & 1.87\\
        B+W             & 0.253 & -0.022 & 2.03\\
        \hline
    \end{tabular}
    %}
    \caption{Performance of the RF model, calculated on the blind test set, applied to different subsets of features. RF$_{\Phi LAB}$ uses the $32$ features selected by $\Phi$LAB, RND uses a set of $32$ features randomly extracted from the original parameter space (best value over the ten extractions), while B+W uses the best $10$ features plus the $22$ excluded by $\Phi$LAB. In all such four parameter spaces both spectroscopic and photometric redshifts were excluded.}
    \label{tab:Philab_Test}
\end{table}

\subsection{Completeness analysis of the training set}
In order to investigate the complexity and completeness of the dataset, we performed three experiments using, as training sets, the full data and two randomly extracted samples from the original training set, consisting of $36,000$ and $100,000$ objects, respectively. We used the $32$ features selected by $\Phi$LAB for these experiments.
As shown in Table~\ref{tab:size_RF}, the RF performance, always calculated on the same blind test set ($241,472$ objects), worsens less than that for MLPQNA with the shrinking of the training set size (see Table~\ref{tab:size_MLPQNA}). Therefore, we use the full amount of data available in the training set in all further experiments.
\begin{table}

\centering
\begin{tabular}{|c | c | c | c |}
\hline
{\bf{Number of}} & \multirow{2}{*}{\bf RMSE} & \multirow{2}{*}{\bf Median} &  \multirow{2}{*}{$\boldsymbol{\eta}$} \\
{\bf{training objects}} & & & \\
\hline \rule{0pt}{3ex}%  EXTRA vertical height
36,000 & 0.278 & -0.022 & 1.99 \\
100,000 & 0.265 & -0.022 & 1.97 \\
362,208 & 0.252 & -0.021 & 2.03  \\
\hline
\end{tabular}
\caption{RF performance against training set size variation. As features we used the best $32$ found by the $\Phi$LAB method and as target the given SFRs. The statistics is calculated on the blind test set.}
\label{tab:size_RF}
\end{table}

\begin{table}

\centering
\begin{tabular}{|c | c | c | c |}
\hline
{\bf{Number of}} & \multirow{2}{*}{\bf RMSE} & \multirow{2}{*}{\bf Median} &  \multirow{2}{*}{$\boldsymbol{\eta}$} \\
{\bf{training objects}} & & & \\
\hline \rule{0pt}{3ex}%  EXTRA vertical height
36,000 & 0.337 & -0.015 & 1.53 \\
100,000 & 0.281 & -0.017 & 1.62 \\
362,208 & 0.248 & -0.017 & 1.99  \\
\hline
\end{tabular}
\caption{MLPQNA performance against training set size variation. As features we used the best $32$ found by the $\Phi$LAB method and as target the given SFRs. The statistics is calculated on the blind test set.}
\label{tab:size_MLPQNA}
\end{table}

\subsection{Redshifts and analysis of dependence from photo-z accuracy}
Looking at the feature importance ranking computed by $\Phi$LAB in Fig.~\ref{fig:featImp}, as it could be expected, the spectroscopic redshifts ($z_{spec}$) carries crucial information to estimate the SFRs. 
Due to the intrinsic uncertainty carried by photometric redshifts, this feature (label $photoz$), has a lower rank ($11th$ out of $56$) and does not seem to carry any particular information contribution to boost the prediction performance. Even if the photoz does not improve the accuracy of the SFR estimation, the presence of both features within the parameter space selected by $\Phi$LAB can be justified by considering that photoz is seen as a noisy version of the more accurate $z\_{spec}$.

In order to evaluate the single contribution of both types of redshift, we performed a set of experiments, reported in Table~\ref{tab:AllFeatures}, by imposing, respectively, a parameter space composed by all $54$ photometric features available without any redshift (experiment PHOT), and the same parameter space in which we alternately added the $z\_{spec}$ (experiment ZSPEC) and $photoz$ (experiment ZPHOT). As it can be seen by looking at the statistical results of Table~\ref{tab:AllFeatures}, the inclusion of $z\_{spec}$ obtains, as expected, better performances, while the presence of $photoz$ seems to be negligible in terms of prediction improvement. However, although the $z_{spec}$ appears as a relevant feature, we dropped it from the used parameter space, since we were interested in predicting SFR via photometric information only.
%\resizebox{\columnwidth}{!}{\centering
\begin{table}
    \centering
    
\resizebox{\columnwidth}{!}{
\begin{tabular}{| c |c | c | c | c |}
        \hline
        \bf ID & \bf features & {\bf RMSE} & {\bf Median}  & {$\boldsymbol{\eta}$} \\
        \hline              
        \rule{0pt}{3ex}%  EXTRA vertical height
        PHOT   & 54 & 0.252 & -0.021 & 1.99\\
        ZSPEC  & 55 & 0.232 & -0.018 & 2.00\\
        ZPHOT  & 55 & 0.252 & -0.021 & 2.18\\
        \hline
    \end{tabular}
    }
    \caption{RF performance over the full set of features. The experiment named PHOT (which contains only magnitudes and Colours) is performed using all the 54 photometric features (i.e. colours and magnitudes); ZSPEC and ZPHOT are two additional experiments, performed by adding to the M+C parameter space, respectively, the spectroscopic and photometric redshift.}
    \label{tab:AllFeatures}
\end{table}
%}

To further verify the Feature selection made by $\Phi$LAB, we repeated the experiments outlined in Table~\ref{tab:AllFeatures} only using the $32$ all-relevant features selected by $\Phi$LAB.
\begin{itemize}
\item RF$_{\Phi LAB}$: experiment using the features identified by $\Phi$LAB (excluding both types of redshift);
\item RF$_{\Phi LAB}+z_{spec}$: experiment with the features identified by $\Phi$LAB including the spectroscopic redshift; 
\item RF$_{\Phi LAB}+z_{phot}$: experiment with the features identified by $\Phi$LAB including the photometric redshift.
\end{itemize}

\begin{table}
\centering

\setlength\tabcolsep{3pt}
\begin{tabular}{|c | c | c | c | c |}
\hline   
\bf ID & \bf features & {\bf RMSE} & {\bf Median} & {$\boldsymbol{\eta}$} \\
\hline 
\rule{0pt}{3ex}%  EXTRA vertical height
 RF$_{\Phi LAB}$ & 32 & 0.252 & -0.021 & 2.03\\
 RF$_{\Phi LAB}+z_{spec}$ & 33 & 0.233 & -0.017 & 2.24\\
 RF$_{\Phi LAB}+z_{phot}$ & 33 & 0.252 & -0.021 & 2.04\\
 \hline
 \end{tabular} 
 \caption{Prediction results of the RF model applied on the blind test set, obtained, respectively, on the parameter space selected by $\Phi$LAB (ID label \textit{RF$_{\Phi LAB}$}) and with the addition of the feature $z_{spec}$ (i.e. spectroscopic redshifts, ID label \textit{RF$_{\Phi LAB}+z_{spec}$}) or $photoz$ (i.e. photometric redshifts, ID label \textit{RF$_{\Phi LAB}+z_{phot}$}).}
 \label{tab:redshifts}
\end{table}

These experiments were performed only with the RF, by excluding MLPQNA due to the much longer training time of this model, assuming also a very similar effect of such additional features on both regression models, by considering our previous analysis done on the $\Phi$LAB feature selection (see Sec. \ref{sec:fs}).

The results, summarised in Table~\ref{tab:redshifts}, confirm that the spectroscopic redshifts bring a higher contribution than the photometric redshifts to estimate SFRs. However, since the two redshifts should in principle represent the same information, we expect that sufficiently accurate photometric redshifts could replace the spectroscopic information, and that any residual prediction error would be dominated by other sources of noise.
Therefore, to get an estimate of how accurate photometric redshifts need to be to obtain a SFR prediction with the same accuracy as reached by including spectroscopic redshifts, we decided to proceed through the following steps: 
\begin{itemize}
%\item estimate the accuracy of the available redshifts;
\item identification of the distribution that fits the $\Delta z_{norm}$ distribution, where $\Delta z_{norm} = (z_{spec} - photoz) / (1 + z_{spec})$;
\item simulation of several $\Delta z_{norm}$ distributions of the same shape, but with different accuracy;
\item application of the different $\Delta z_{norm}$ to the $z_{spec}$ in order to simulate $photoz$ with increasing accuracy;
\item testing the SFR estimation using simulated $photoz$.
\end{itemize}

\begin{figure}
\centering
\includegraphics[width=\columnwidth]{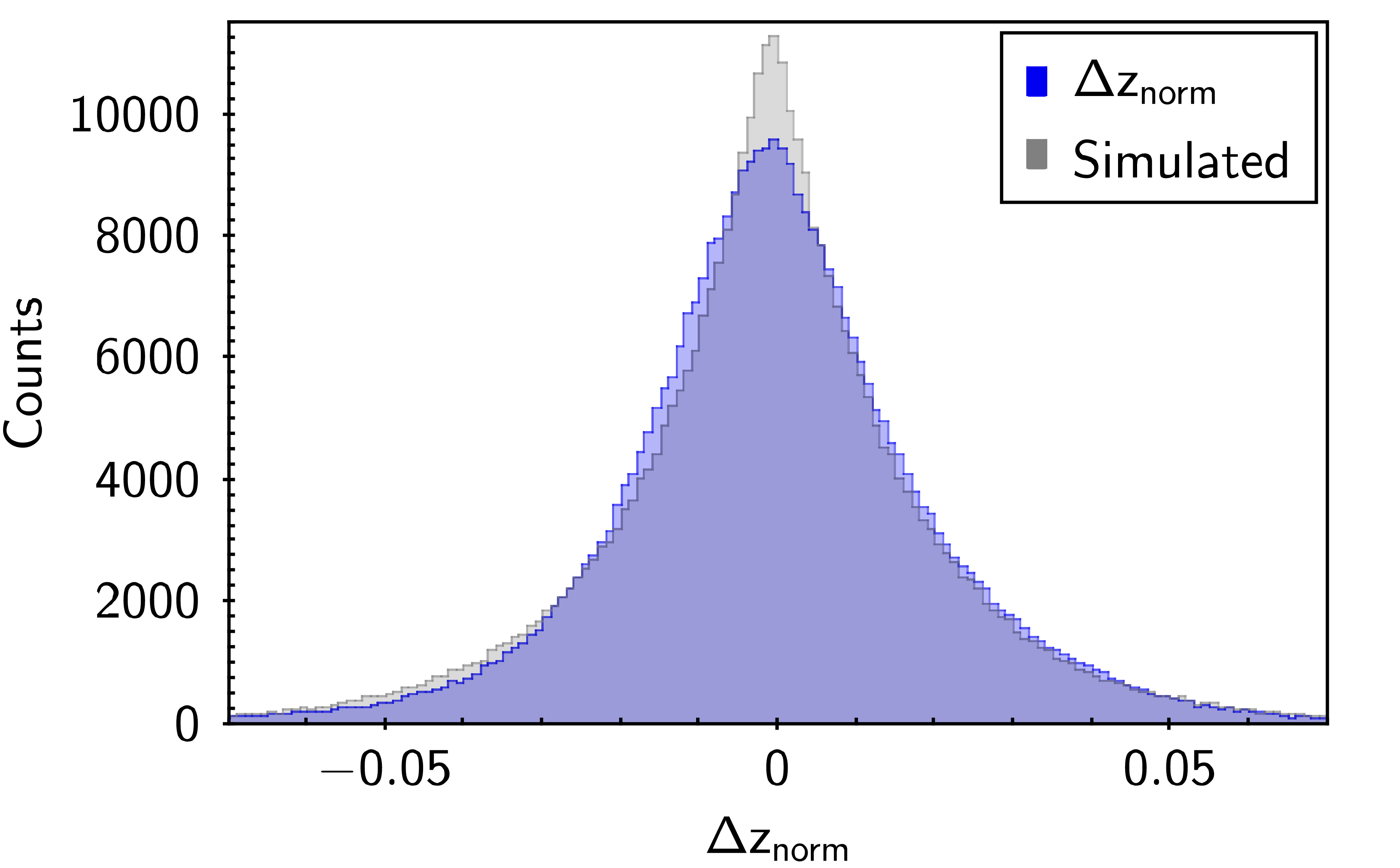}
\caption{Distribution of redshift residuals $\Delta z_{norm}$ (colored in blue) with the superimposed best fitting Laplacian distribution (coloured in gray).}\label{fig:Deltadist}
\end{figure}

We started by calculating the $\Delta z_{norm}$ distribution of the $photoz$ used for the  RF$_{\Phi LAB}+z_{phot}$ experiment, obtaining a distribution with a bias of $-0.00079$ and a $\sigma$ of $0.022$.
We then estimated through a Kolmogorov-Smirnov test \citep{KST} the distribution that best fits the $\Delta z_{norm}$ distribution of the $photoz$. We tried to fit the data with all the continous distributions implemented in the \textit{scipy.stats} module\footnote{\url{https://docs.scipy.org/doc/scipy/reference/stats.html}}. 
A Laplacian distribution with a standard deviation of $0.015$ and a bias of $0.0077$ was found to be the best fit, see Fig.~\ref{fig:Deltadist}. 
This  distribution was then used to generate random noise that we added to the original $z_{spec}$ distribution in order to simulate the $photoz$'s (and thus its measurement error). The process of noise generation and addition was repeated ten times in order to compare the resulting SFR estimation statistics and to make sure that the correlation between the simulated error and the corresponding statistics was consistent. Afterwards, we repeated the  RF$_{\Phi LAB}+z_{phot}$ experiment using this new photometric redshift distributions finding an average RMSE variation along the ten extractions of $\sim 0.001$.

As reported in the first row of Table~\ref{tab:photoz}, the statistical performance is very similar to the  RF$_{\Phi LAB}+z_{phot}$ experiment (Table~\ref{tab:redshifts}), thus proving that our simulation is able to reproduce the behaviour of photometric redshifts (the slight difference in performance may be due to the presence of systematic errors, ignored by the simulation). We then proceeded to an iterative decrease of the $\sigma$ of the Laplacian distribution, in order to simulate 
%a variable 
an increasing quality of $photoz$ estimations; at each step we repeated (ten times) the RF$_{\Phi LAB}+z_{phot}$ experiment with the new distribution of $photoz$. The results are reported in Table~\ref{tab:photoz} and show that, in order to obtain an efficiency comparable with the one obtained using the spectroscopic redshifts, an accuracy of at least $\sigma=0.005$ is required for the $photoz$ estimation. 
We want to underline that this is simply an indication of the photometric redshift accuracy required to become indistinguishable from the SFR prediction accuracy reached with spectroscopic redshifts. 
This standard deviation value is lower than what can be found in literature; see for instance \citealt{Brescia2014} ($\sigma=0.028$ in the range $0 < z_{spec} \le 1$) or \citealt{Laurino} ($\sigma=0.015$ in the range $0 < z_{spec} \le 0.65$) or \citealt{Ball} ($\sigma=0.021$ in the range $0 < z_{spec} \le 0.5$), motivated by the smallest redshift range considered in this particular case ($0 < z_{spec} \le 0.33$).

\begin{table}

\centering
\begin{tabular}{| c | c | c | c |}
\hline
\textbf{redshift used} & \textbf{RMSE} & \textbf{Median} &  {$\boldsymbol{\eta}$} \\
\hline
\rule{0pt}{3ex}%  EXTRA vertical height
$\sigma=0.022$ & 0.249 & -0.019 & 2.08\\
$\sigma=0.015$ & 0.244 & -0.019 & 2.11\\
$\sigma=0.007$ & 0.238 & -0.018 & 2.18\\
$\sigma=0.005$ & 0.236 & -0.018 & 2.21\\
RF$_{\Phi LAB}+z_{spec}$ & 0.233 & -0.017 & 2.24\\
\hline
\end{tabular}
\caption{Photometric redshift accuracy estimation experiments. The first four experiments are referred to the SFR RF$_{\Phi LAB}+z_{phot}$ estimations varying the $photoz$ measurement precision. While in the last one the photometric redshifts were replaced by spectroscopic redshifts.} 
\label{tab:photoz}
\end{table}

\subsection{Catastrophic Outliers}\label{sec:c_outliers}
\begin{figure*}
    \centering
    \includegraphics[width=0.45\textwidth]{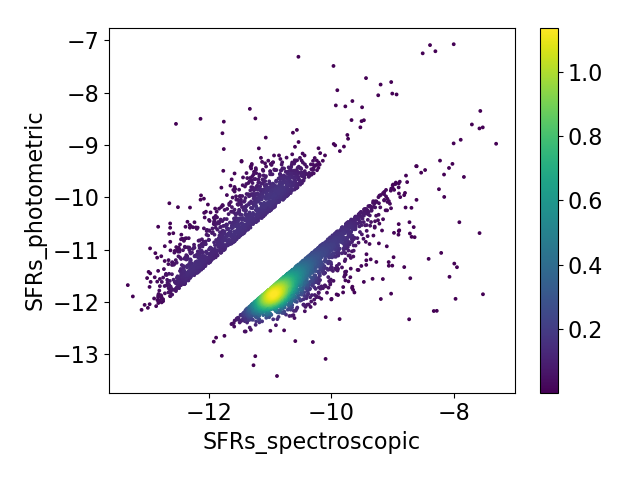}(a) \quad
    \includegraphics[width=0.45\textwidth]{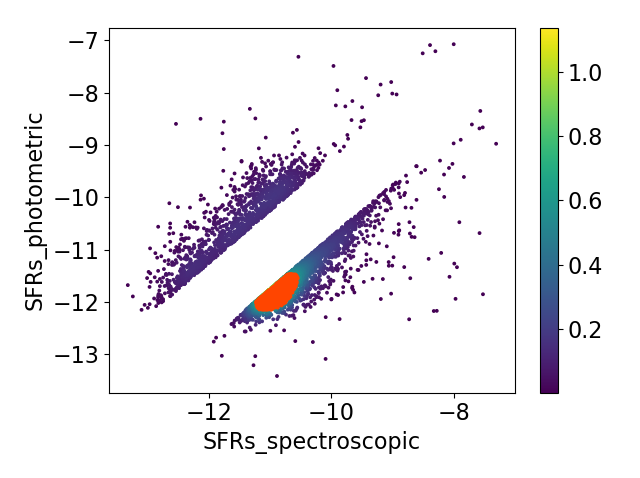}(b) \\
    \includegraphics[width=0.45\textwidth]{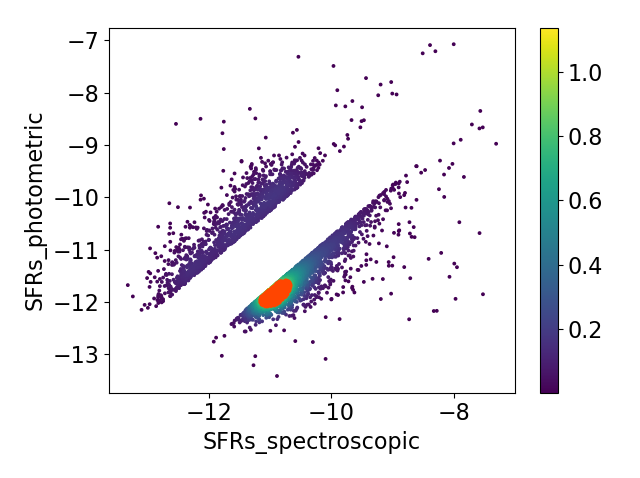}(c) \quad
    \includegraphics[width=0.45\textwidth]{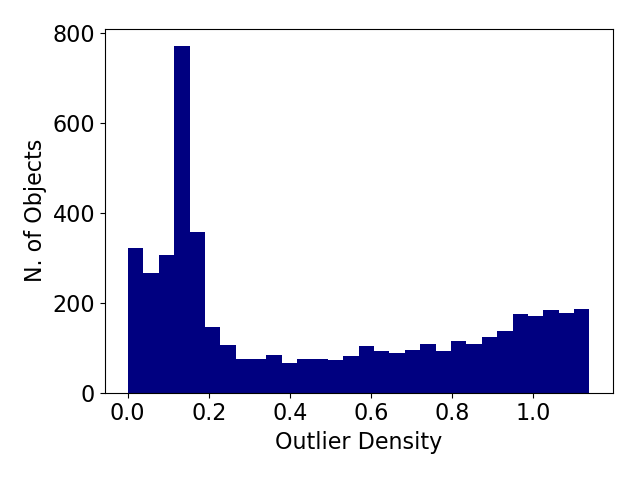}(d)
    
    \caption{The scatter plot in the top left corner (a) shows the distribution of outliers in the $SFRs_\text{spectroscopic}$ VS $SFRs_\text{photometric}$ space with a superimposed density map, while the diagrams int the top right (b) and bottom left (c) corners show highlighted in orange all the objects with a density, respectively, six and eight times higher than the average point density. The histogram in the bottom right corner (d) shows the outliers density distribution.}
    \label{fig:Over-density}
\end{figure*}

As already mentioned, due to the higher accuracy, we decided to use the MLPQNA model to create our SFRs catalogue (see Sec. \ref{sec:fs_results}), so in order to detect possible issues with the model and gain insights into the nature of the physical problem, we analysed the nature of the catastrophic outliers (i.e. those objects whose SFR prediction error resulted higher than $3\sigma$) distribution relative to the MLPQNA$_{\Phi LAB}$ experiment.
In Fig.~\ref{fig:Over-density} it is shown the distribution of catastrophic outliers in the $SFRs_\text{spectroscopic}$ VS $SFRs_\text{photometric}$ space, resulting from the MLPQNA$_{\Phi LAB}$ experiment reported in Table~\ref{tab:mlpqnavsrf}. 
We estimated the pixel density through a kernel density estimation method \citep{scott1992} and coloured the pixels on the basis of their density. 
As shown in the scatter plot of Fig.~\ref{fig:Over-density}(a), most of the point are clustered in a small region (highlighted in yellow) hereafter called the \emph{over-density region} (that is confirmed also using the Random Forest results). In order to understand why these objects are outliers, we selected all the objects belonging to the over-density region through cuts in their local density. 
The scatter plots of Fig.~\ref{fig:Over-density}(b) and Fig.~\ref{fig:Over-density}(c) show highlighted in orange all the objects with a density, respectively, six and eight times higher than the average point density. 
Depending on the cuts, the over-density region contains $1,877$ objects (six times the average density) or $1,277$ objects (eight times the average density) out of the total number of $4,840$ objects classified as catastrophic outliers. 
We then investigated the possibility that these objects could form a cluster in some bi-dimensional projections of the parameter space. We tried all the possible magnitudes, colours, and redshifts combinations without finding any obvious clustering (some of these combinations are shown in Appendix \ref{sec:projections}). We also checked whether the group could correlate with a specific (high) error measure associated to any of the used features, but no any evident correlations were found.
The nature of the objects in the over-density region is still under further investigation.

\subsection{Comparison with a recent work}\label{sec:comparison}
In order to compare our regression models with the k-NN used by \cite{Stensbo} and their feature selection, we performed an experiment using the full training set and the set of $8$ features found by \cite{Stensbo}. In Table \ref{tab:comparison} we present the statistical results, which show a comparable performance among the three methods, although with a lower RMSE obtained by RF and MLPQNA.
Using the features found by $\Phi$LAB, the RF and MLPQNA can achieve even better performance, as shown in table~\ref{tab:mlpqnavsrf}. This is not surprising as k-NN is much more sensitive to the dimensionality of the parameter space (the so-called ``curse of dimensionality'') than other two models. These latter can, therefore, take advantage of the information carried by a larger number of features than a k-NN model.

\begin{table}
    \centering
    \begin{tabular}{|c | c | c | c |}
    \hline
    \textbf{Model} & \textbf{RMSE} & \textbf{Median} &  {$\boldsymbol{\eta}$} \\
    \hline
    RF     & 0.264 & -0.020 &  1.86\\
    k-NN   & 0.274 & 0.013  &  1.85 \\
    MLPQNA & 0.265 & -0.021 &  1.85 \\
    \hline
    \end{tabular}
    \caption{Comparison between our RF and MLPQNA against \citep{Stensbo} k-NN using the full train set and the best 8 features found by Stensbo-Smidt.}
    \label{tab:comparison}
\end{table}

\section{Discussion and Conclusions}\label{sec:concl}
In this work, based on our preliminary analysis of the problem presented at the ESANN-2018 conference \citep{DelliVeneri}, we estimated star formation rates for a large subset of the Sloan Digital Sky Survey DR-7 and produced a catalogue of SFRs derived using photometric features only (magnitudes and colours) and the MLPQNA machine learning model (see Appendix \ref{catalogue}) trained on a knowledge base of spectroscopically determined SFRs. 
By looking at Fig.~\ref{fig:specvspredict} and the statistics in Table~\ref{tab:mlpqnavsrf}, the regression results appear very promising. This is particularly true, considering that the dynamical range of SFR is between $-12$ and $-7$, and also that we have $\sim 5,000$ outliers out of the $242,000$ objects of the blind test set and, finally, taking into account the low percentage of outliers ($\sim 2$\%). However, from the results obtained by varying the size of the model training set (Tables \ref{tab:size_RF} and \ref{tab:size_MLPQNA}), we think that a larger knowledge base of SFRs would further improve the performances. 
 
Furthermore, the residual scatter is likely to be an artefact of the photometry. The figure $5a$ in \cite{Stensbo} shows a scatter plot for the predictions obtained with SED fitting. It appears qualitatively similar to the current work and could suggest that there is a more fundamental limit to the accuracy we can expect from optical photometry only.
This is not an obvious issue; for example in \cite{Brescia2014} it was demonstrated, in the case of estimation of photometric redshifts, that the model performance, over a certain amount of data, does not scale with the size of the training set. 

By considering the median estimator, in all our experiments its values are always negative. This is a consequence of the presence of the over-density described in Sec.~\ref{sec:c_outliers} and shown in Fig.~\ref{fig:specvspredict}. We intend to perform a deeper investigation on such objects, which will focus on the characterisation of objects in the overdensity region in terms of their spectroscopic, morphological and evolutionary properties.

By applying the $\Phi$LAB method, we found the all-relevant set of features and were able to discard almost half of the initial set of features without any loss in precision over the full set, but with a great gain in computing time. We tested the $\Phi$LAB method several times, confirming the reliability of its feature selection.

Since in future surveys it is likely that no large spectroscopic samples will be available, we run a simulation to find the minimum accuracy required for photometric redshifts in order to effectively replace spectroscopic estimates, finding that SFRs can be predicted with the same accuracy under the condition to provide photo-z with an error smaller than $0.005$ (See Table~\ref{tab:photoz}).

From our results on the Sloan Digital Sky Survey DR7, we think that our machine learning methods could be applied to other surveys to reliably calculate SFRs. On this note we intend to expand our photometric knowledge base to the UV, X-ray and infrared in order to:
\begin{itemize}
    \item use the full spectrum to identify and constrain outliers and potential issues in the methods (i.e. AGN selection through X-ray photometry);
    \item incorporate the UV and infrared information to derive SFRs.
\end{itemize}
Moreover we intend to apply our methods to to derive photometric SFRs from the ESO-KiDS-DR4 (Kuijken et al. in prep.).
We wish to conclude by saying that the natural evolution of this work will be to expand our knowledge base above $z_{spec} = 0.33$. In this case, on one hand, redshifts would have a bigger impact on galaxy emission and thus magnitudes; on the other, we should be able to produce high quality SFRs for larger samples of objects. 

\section{acknowledgements}\label{sec:acknowlg}
The authors would like to thank the anonymous referee for extremely valuable comments and suggestions. A special thank goes to Kristoffer Stensbo-Smidt for all the work and the precious suggestions he put in reviewing and commenting this work.
MB acknowledges the \textit{INAF PRIN-SKA 2017 program 1.05.01.88.04} and the funding from \textit{MIUR Premiale 2016: MITIC}.
GL and MDV acknowledge the UE funded Marie Curie \textit{ITN SUNDIAL} which partially supported the present work. 
SC acknowledges support from the project ``\textit{Quasars at high redshift: physics and Cosmology}'' financed by the ASI/INAF agreement 2017-14-H.0.
Topcat has been used for this work \citep{Taylor2005}. C$^3$ has been used for catalogue cross-matching \citep{riccio17}. DAMEWARE has been used for machine learning experiments \citep{Brescia2014783}.

\bibliographystyle{mnras}
\interlinepenalty=1000
\bibliography{main}
\clearpage

\onecolumn
\appendix
\section{\mathinhead{\Phi}{Phi}LAB Method}\label{sec:philabmethod}
$\Phi$LAB is based on the combination of two components: \textit{shadow features} and \textit{Na\"{\i}ve LASSO} statistics.
The term \textit{shadow features} arises from the idea to extend the given parameter space with artificial features \citep{Kursa2010}. Given a dataset of $N$ samples, represented through a $D$-dimensional parameter space, we introduce a shadow feature for each real one, by randomly shuffling its values among the $N$ samples, thus doubling the original parameter space.
Shadow features are, thus, random versions of the real ones and their importance percentage can be used as a threshold for when a real feature is containing actual information. Such a threshold is important since feature selection methods only provide a ranking of the features, never an absolute important/not important decision.
The second component of $\Phi$LAB is based on the \textit{Na\"{\i}ve LASSO} statistics. The LASSO \citep[Least Absolute Shrinkage and Selection,][]{Tibshirani2012} performs both a variable selection and a regularisation of a ridge regression (i.e. a shrinking of large regression coefficients to avoid overfitting), enhancing the prediction accuracy of the statistical model. The regularisation is a typical process exploited within ML, based on the addition of a functional term to a loss function (e.g. a penalty term). LASSO performs the so-called $L_1$ regularisation (i.e. based on the standard $L_1$ norm), which has the effect of \emph{sparsifying} the weights of the features, effectively turning off the least informative features. In particular, we included two \textit{Na\"{\i}ve LASSO} techniques in $\Phi$LAB.
One is the A-LASSO (Alternate-LASSO; \citealt{Hara2016}), able to find all weakly relevant features that could be removed from the standard LASSO solution. Such method calculates a list of features alternate to those selected by the standard LASSO, each one associated with a calculated score, reflecting the performance degradation from the optimal solution. In $\Phi$LAB, we select only the alternate features that achieve the lowest score difference from the best features, trying to reach the best trade-off between feature selection performance and flexibility in the analysis of the parameter space.
Such alternate features smoothly degrade the solution score, but may potentially infer more flexibility, by relaxing the intrinsic stiffness of the best solution system.
The second version of the standard LASSO is E-LASSO (Enumerate-LASSO; \citealt{Hara2017}), which enumerates a series of different feature subsets, considered as solutions with a decreasing level of approximation. The main concept behind is that an optimal solution to a mathematical model is not necessarily the best solution to any physical problem. Therefore, by enumerating a variety of potential solutions, there is a chance to obtain better solutions for the problem domain task. For instance, Hara and Maehara demonstrate that E-LASSO solutions are good approximations to the optimal solution, by also improving the flexibility for the selection of the parameter space, covering a wide spectrum of variations within the problem domain (i.e. by helping to find the all-relevant set of features).
The shadow features and \textit{Na\"{\i}ve LASSO} are then combined by selecting the candidate weak relevant features through the shadow feature noise threshold and by extracting the final set of weak relevant features through a filtering process, based on the A-LASSO and confirmed by E-LASSO.
To summarise, we find the list of candidate features through the shadow features technique and then we use the LASSO operator to explore the parameter space and verify the effective contribution carried by those features considered as weak relevant to the solution of the problem. The loss function based on L1 regularisation is crucial to quantify the degradation of performance when other features subsets are replacing the best one, by also automatically identifying the exact redundancy of some features that the shadow features technique is unable to disentangle in terms of individual importance.

\begin{figure*}
\centering
\includegraphics[width=\textwidth,height=.9\textheight,keepaspectratio]{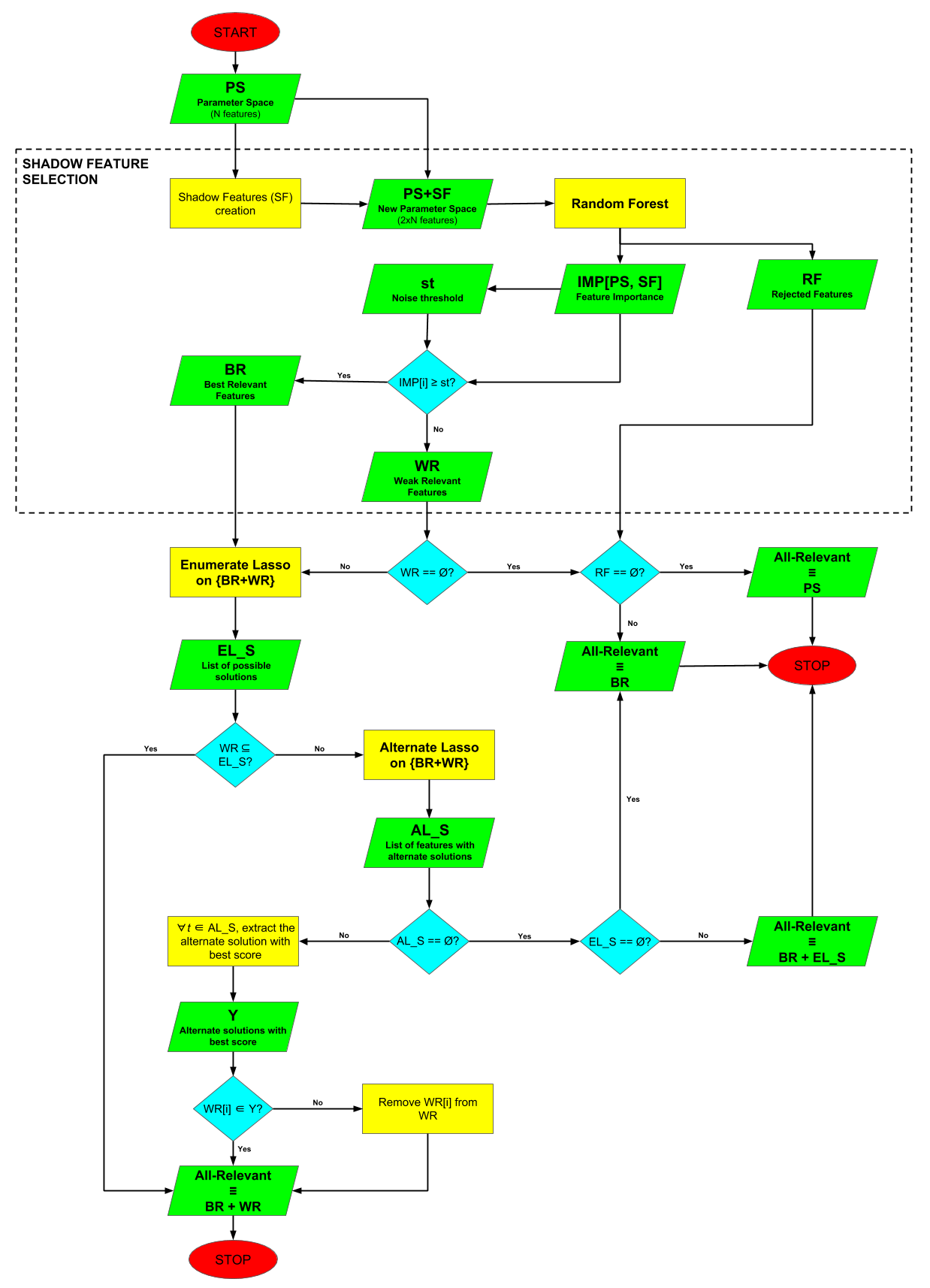}
\caption{$\Phi$LAB workflow}
\label{fig:philab}
\end{figure*}

The pseudo-code of the features selection method can be summarised by the following steps (see also Fig. \ref{fig:philab}):

\begin{enumerate}
\item Let the set \{x$_1$, x$_2$, ..., x$_D$\} be the initial complete parameter space composed by D real features; 
\item Apply the Shadow Feature Selection (SFS method) and produce the following items:
\begin{itemize}
\item[-] SF={x$_{s_1}$\dots x$_{s_D}$}, the list of shadow features, obtained by randomly shuffling the values of real features;
\item[-] max(IMP[parameter space, SF]) $\forall x \in$ parameter space \& $\forall x_s \in SF$, the importance list of all 2D features, original and shadows;
\item[-] st: noise threshold, defined as the max\{IMP[SF], $\forall x_s \in$ SF\};
\item[-] BR=\{$x \in$ parameter space with IMP[x] $\ge$ st\}, the set of best relevant real features;
\item[-] RF=\{x $\in$ parameter space, rejected by the Shadow Feature Selection\}, the set of excluded real features, i.e. not relevant;
\item[-] WR=\{x $\in$ parameter space with IMP[x] $<$ st\}, the set of weak relevant real features;
\end{itemize}
\item At this stage the complete parameter space is now split into $BR$, $WR$ and $RF$. Now we consider the reduced parameter space, space$_{red}$= \{BR+WR\}, obtained by excluding the rejected features. In principle it may correspond to the original parameter space if there is no rejections by the SFS;
\begin{itemize}
\item[(A)] If RF==$\emptyset$ \&\& WR==$\emptyset$, the SFS method confirmed all real features as high relevant, therefore return ALL-RELEVANT (parameter space), i.e. the full parameter space as the optimised parameter space and EXIT.
\item[(B)] If RF$\ne\emptyset$ \&\& WR==$\emptyset$, the SFS method rejected some features and confirmed others as high relevant, therefore return ALL-RELEVANT (BR) as the optimised parameter space and EXIT.
\item[(C)] If WR$\ne\emptyset$, regardless some rejections, SFS confirmed the presence of some weak relevant features that must be evaluated by LASSO methods, therefore go to step (iv);
\end{itemize}
\item Apply E-LASSO method on the space$_{red}$= \{BR+WR\} producing:
\begin{itemize}
\item[-] EL\_S: a list of M subsets of features, considered as possible solutions, ordered by decreasing score;
\item[(A)] If WR $\subseteq$ EL\_S, then all weak relevant features are possible solutions, therefore return ALL-RELEVANT(BR+WR) as the optimised parameter space and EXIT.
\item[(B)] Else go to step (v); 
\end{itemize}
\item Apply A-LASSO method on the space$_{red}$= \{BR+WR\}(set of candidate features) producing:
\begin{itemize}
\item[-] AL\_S, a set of T features, each one with a corresponding list of features List(t) considered as alternate solutions with a certain score;
\item[(A)] if AL\_S ==$\emptyset$ then no alternate solutions exist, therefore:
\begin{itemize}
\item[(A.1)] If EL\_S==$\emptyset$ then return ALL-RELEVANT(BR) as the optimised parameter space and EXIT.
\item[(A.2)] Else if EL\_S$\ne\emptyset$ then return ALL-RELEVANT(BR+EL\_S) as the optimised parameter space and EXIT.
\end{itemize}
\item[(B)] Else extract $\forall t \in T$ the alternate solution with Score(as) = min\{Score(y), $\forall y \in$ List(t)\};
\item[(C)] go to step (vi).
\end{itemize}
\item For each $x \in $WR:
\begin{itemize}
\item[(A)] If x is alternate solution of at least one feature $t \in T$, with [t $\in$ BR || t $\in$ EL\_S], then retain x within WR set;
\item[(B)] Else reject x (by removing x from WR);
\end{itemize}
\item Return ALL-RELEVANT(BR+WR) as the final optimised parameter space and EXIT.
\end{enumerate}

\section{Catalogue} \label{catalogue}
We produced a SFR catalogue containing SFRs for $27,513,324$ galaxies of the SDSS-DR7, which is accessible through the Vizier facility at the following link \url{ftp://cdsarc.u-strasbg.fr/pub/cats/J/MNRAS/486/1377}. To produce the catalogue, we started by querying the \textit{Galaxy} View\footnote{\url{http://skyserver.sdss.org/dr7/en/help/browser/browser.asp?n=Galaxy&t=U}} of the SDSS-DR7 for all the needed photometric features of galaxies with a "good" photometry (see PhotoFlags) and containing no \textit{Missing Values}. We then applied the magnitudes cuts of our knowledge base (in order to keep the photometric features within the ranges of our knowledge base) and cross-matched the resulted dataset with the \textit{photoz} catalogue derived by \cite{Brescia2014}, in order to use them as a quality flag. The final catalogue contains the following columns: 
\begin{itemize}
\item Identifiers: \textit{dr9objid}, \textit{objid}, \textit{ra}, \textit{dec}, i.e. respectively the object identifier in the SDSS DR9 and DR7 and their ascension and declination coordinates;
\item Quality Flags: $photoz$ and \textit{Quality\_Flag}, i.e. the \textit{photometric redshifts} measured by  \cite{Brescia2014} and the associated flag. The \textit{Quality\_Flag} can assume three values $1$, $2$ and $3$; $1$ stands for the best photo-z accuracy, $2$ and $3$ for decreasing accuracy;
\item SFR: Star Formation Rate computed by the MLPQNA model with the $32$ best features selected by the $\phi$LAB method (excluding redshifts).
\end{itemize}
In order to select only SFRs with high quality (i.e. only select sources inside the training set parameter space constrains), the user should impose $photoz \leqslant 0.33$  and $Quality\_Flag = 1$. This is due by considering that in our knowledge base there are only objects with spectroscopic redshift less than $0.33$, thus we are able to predict SFRs only for objects within such redshift range. These constraints will select $\sim 6.6$ million objects.
Since we do not have any spectroscopic redshifts for the catalogue objects, we must use photometric redshifts (where available) to perform these cuts. Nevertheless using photometric redshifts instead of spectroscopic ones, may introduce some contamination in the catalogue, i.e. a source may be inside the $photoz \leqslant 0.33$ cut when in reality it has a spectroscopic redshift higher than $0.33$. To estimate the number of such contaminants, we verify that among the $871,784$ objects with $photoz \leqslant 0.33$  and a spectroscopic redshift only  $\sim1.33\%$ resulted to have a true redshift higher that $0.33$.

\section{BI-DIMENSIONAL PROJECTIONS TO ISOLATE THE Over-density REGION}\label{sec:projections}
In this section we show some examples of bi-dimensional projections in the parameter space, among the most relevant features, done in order to isolate the objects in the Over-density Region. As stated in Sec. \ref{sec:c_outliers}, no projections were found able to achieve such separation.
\begin{figure*}
    \centering
    \includegraphics[width=0.48\textwidth]{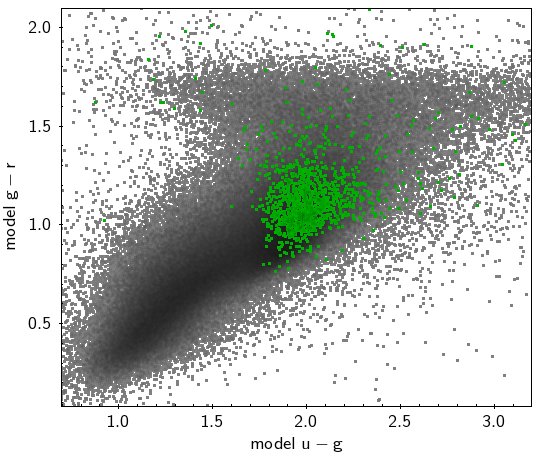}
    \includegraphics[width=0.48\textwidth]{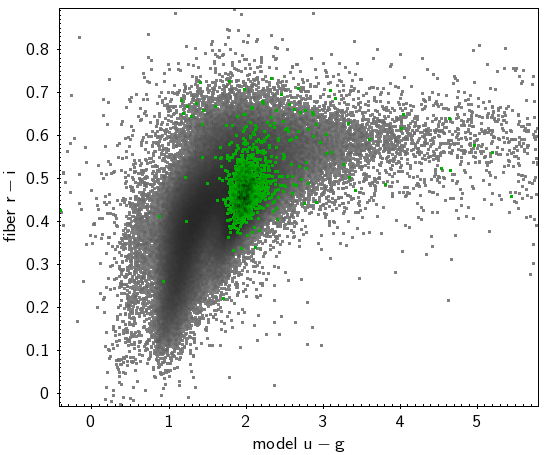}\\
    \includegraphics[width=0.48\textwidth]{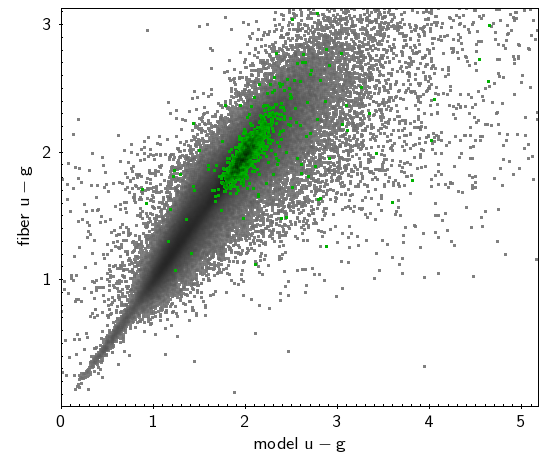} 
    \includegraphics[width=0.48\textwidth]{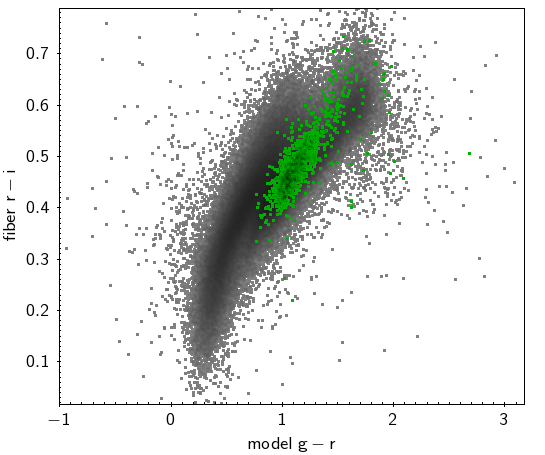}\\
    \includegraphics[width=0.48\textwidth]{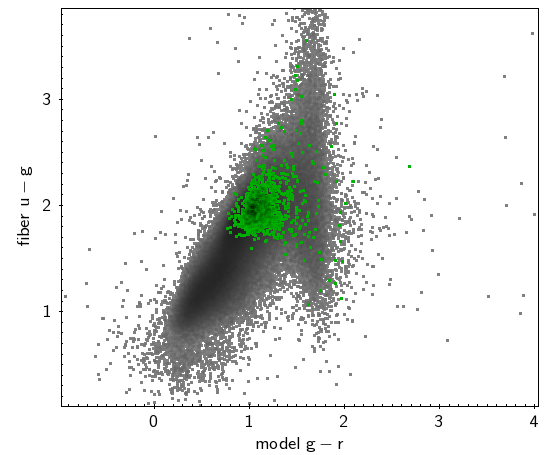}
    \includegraphics[width=0.48\textwidth]{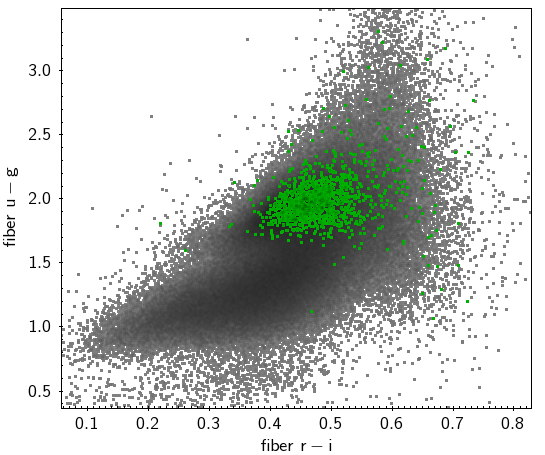} 
    \caption{Some examples of bi-dimensional projections of the parameter space, done in order to isolate the objects of the Over-density Region shown in Fig.~\ref{fig:Over-density}. In particular, all the combinations of the most relevant colours are shown. The objects belonging to the Over-density Region are highlighted in green colour.}
\end{figure*}

\section{Example of queries used to obtain galaxies from the SDSS-DR7}

\begin{Verbatim}
SELECT
    p.objid, p.ra, p.dec, 
    p.modelMag_u,  p.modelMag_g, p.modelMag_r,  p.modelMag_i,  p.modelMag_z, 
    p.devMag_u,    p.devMag_g,   p.devMag_r,    p.devMag_i,    p.devMag_z, 
    p.expMag_u,    p.expMag_g,   p.expMag_r,    p.expMag_i,    p.expMag_z, 
    p.petroMag_u,  p.petroMag_g, p.petroMag_r,  p.petroMag_i,  p.petroMag_z, 
    p.fiberMag_u,  p.fiberMag_g, p.fiberMag_r,  p.fiberMag_i,  p.fiberMag_z, 
    p.psfMag_u,    p.psfMag_g,   p.psfMag_r,    p.psfMag_i,    p.psfMag_z, 
    q.objid as dr9objid 
INTO
    mydb.p75p90 
FROM 
    Galaxy as p, 
    dr9.PhotoPrimaryDR7 as s, 
    dr9.Galaxy as q
WHERE
    p.mode = 1                  AND
    p.dec >= 75 AND p.dec < 90  AND
    s.dr7objid   =  p.objid     AND
    s.dr8objid   =  q.objid     AND 
    p.modelMag_u > -9999        AND        p.modelMag_g > -9999        AND 
    p.modelMag_r > -9999        AND        p.modelMag_i > -9999        AND 
    p.modelMag_z > -9999        AND        p.devMag_u   > -9999        AND 
    p.devMag_g   > -9999        AND        p.devMag_r   > -9999        AND
    p.devMag_i   > -9999        AND        p.devMag_z   > -9999        AND 
    p.expMag_u   > -9999        AND        p.expMag_g   > -9999        AND
    p.expMag_r   > -9999        AND        p.expMag_i   > -9999        AND 
    p.expMag_z   > -9999        AND        p.petroMag_u > -9999        AND
    p.petroMag_g > -9999        AND        p.petroMag_r > -9999        AND 
    p.petroMag_i > -9999        AND        p.petroMag_z > -9999        AND
    p.fiberMag_u > -9999        AND        p.fiberMag_g > -9999        AND 
    p.fiberMag_r > -9999        AND        p.fiberMag_i > -9999        AND
    p.fiberMag_z > -9999        AND        p.psfMag_u   > -9999        AND 
    p.psfMag_g   > -9999        AND        p.psfMag_r   > -9999        AND
    p.psfMag_i   > -9999        AND        p.psfMag_z   > -9999        AND
    dbo.fPhotoFlags('PEAKCENTER')          != 0                        AND
    dbo.fPhotoFlags('NOTCHECKED')          != 0                        AND
    dbo.fPhotoFlags('DEBLEND_NOPEAK')      != 0                        AND
    dbo.fPhotoFlags('PSF_FLUX_INTERP')     != 0                        AND
    dbo.fPhotoFlags('BAD_COUNTS_ERROR')    != 0                        AND
    dbo.fPhotoFlags('INTERP_CENTER')       != 0                           

\end{Verbatim}
% Don't change these lines
\bsp	% typesetting comment
\label{lastpage}
\end{document}